\documentclass[preprint,12pt]{elsarticle}

\usepackage[tight,footnotesize]{subfigure}
\usepackage{amssymb}
\newtheorem{definition}{Definition}[section]
\newtheorem{theorem}{Theorem}[section]
\newtheorem{lemma}{Lemma}[section]

\journal{Elsevier Science}

\begin{document}

\begin{frontmatter}

%% Title, authors and addresses

%% use the tnoteref command within \title for footnotes;
%% use the tnotetext command for the associated footnote;
%% use the fnref command within \author or \address for footnotes;
%% use the fntext command for the associated footnote;
%% use the corref command within \author for corresponding author footnotes;
%% use the cortext command for the associated footnote;
%% use the ead command for the email address,
%% and the form \ead[url] for the home page:
%%
%% \title{Title\tnoteref{label1}}
%% \tnotetext[label1]{}
%% \author{Name\corref{cor1}\fnref{label2}}
%% \ead{email address}
%% \ead[url]{home page}
%% \fntext[label2]{}
%% \cortext[cor1]{}
%% \address{Address\fnref{label3}}
%% \fntext[label3]{}

\title{Probabilistic Performance Analysis of Networks using an Improved Network Service Envelope Approach}

%\title{A New Network Service Envelope for Stochastic Performance Analysis of Networks using Network Calculus}
%% use optional labels to link authors explicitly to addresses:
%%\author{K. Angrishi}
%%\address{T-Systems International GmbH\\
%%20146 Hamburg, Germany\\
%%kishore.angrishi@t-systems.com}

\author{K. Angrishi, U. Killat}
\address{Institute of Communication Networks\\
Hamburg University of Technology\\
21075 Hamburg, Germany\\
\{kishore.angrishi,killat\}@tu-harburg.de}

\begin{abstract}
Stochastic network calculus is an evolving theory which accounts for statistical multiplexing and uses an envelope approach for probabilistic delay and backlog analysis of networks. One of the key ideas of stochastic network calculus is the possibility to describe the service offered at a network node as a stochastic service envelope, which in turn can be used to describe the stochastic service available in a network of nodes and determine end-to-end probabilistic delay and backlog bounds. This paper introduces a new definition of stochastic service envelopes which yields a simple network service envelope and tighter end-to-end performance bounds. It is shown for ($\sigma(\theta), \rho(\theta)$) - constrained traffic model  that the end-to-end performance measures computed using the new stochastic network service envelope are tight in comparison to the ones obtained using the existing start-of-the-art definition of statistical network service envelope and are bounded by ${\cal O}(H \log{H})$, where $H$ is the number of nodes traversed by the arrival traffic.
%is tighter than the bounds determined with the existing state of the art definition of stochastic network service envelope while maintaining the ${\cal O}(H \log{H})$ scaling properties, where $H$ is the number of nodes traversed by the arrival traffic.  The benefits of the new stochastic service envelope is illustrated using ($\sigma(\theta), \rho(\theta)$)- constrained traffic model.
\end{abstract}

\begin{keyword}
Stochastic Network Calculus \sep Network Service Envelope \sep Quality of Service
%% keywords here, in the form: keyword \sep keyword
%% MSC codes here, in the form: \MSC code \sep code
%% or \MSC[2008] code \sep code (2000 is the default)
\end{keyword}

\end{frontmatter}
%%
%% Start line numbering here if you want
%%
% \linenumbers
%% main text
\section{Introduction}
\label{sec:intro}

The convergence of data, voice and video traffic over the Internet has increased the significance of performance analysis of data networks. The critical aspect in the performance analysis of data networks is the efficient modeling of arrival traffic and service available to the arrival traffic in the network. The exactness of the performance measures depends on the accuracy of the models describing the arrival traffic and service available in the network, but with increased mathematical complexity. In most cases, bounds on the performance measure are sufficient for network analysis. Network calculus is one of the popular theories useful for computing worst-case performance bounds in data networks with the help of deterministic envelopes describing the arrival traffic and service available in a network node. The probabilistic version of network calculus is called stochastic network calculus \footnote{The terms statistical network calculus, stochastic network calculus and probabilistic network calculus are used interchangeably in the literature} which retains the envelope approach and many favorable characteristics of (deterministic) network calculus and derives probabilistic performance bounds. The raison d'\^etre of network calculus is the possibility to compute probabilistic bounds on end-to-end performance measures using a network service envelope which describes the service available in a network as a single abstract node. It has been shown in \cite{boudec:2001} that the end-to-end worst case performance bounds obtained by summing the per-node results scale in the order of ${\cal O}(H^2)$ and the bounds computed using network service envelope scale in the order of ${\cal O}(H)$, where $H$ is the number of nodes traversed by the arrival flow. There has been many attempts to achieve a similar linear scaling of end-to-end performance bounds in statistical network calculus, but with little success. We direct the interested readers to \cite{li:2007} for a more elaborate discussion on what makes the probabilistic extension of network calculus so difficult. In \cite{florin:2006}, authors presented a stochastic network service envelope which allows the computation of end-to-end probabilistic performance measures that is shown to be bounded by ${\cal O}(H \log{H})$ for exponential bounded burstiness (EBB) traffic model.

In this paper, we present a different definition for statistical service envelope based on the stochastic service process characterizing the service offered at a network node. The new definition of statistical service envelope allows to compute tighter end-to-end performance measures than the ones obtained using the existing definition of network service envelope from \cite{florin:2006} while still maintaining the ${\cal O}(H \log{H})$ scaling of the end-to-end bounds for ($\sigma(\theta), \rho(\theta)$) - constrained traffic model. Later in the paper, we will use Markov modulated on-off traffic model as in \cite{florin:2006} to demonstrate the tightness of the delay measure computed using the new definition of statistical service envelope.

The rest of the paper is structured as follows: In Section \ref{sec:SNC}, we give an overview of the statistical network calculus and define our notion of statistical service envelope. Then, we use the envelope functions to derive performance bounds on delay, backlog and output burstiness. The scaling properties of the derived end-to-end performance bounds are shown in Section \ref{sec:scale}. In Section \ref{sec:num}, a numerical example using Markov Modulated On-Off traffic is presented for illustration. Brief conclusions are presented in Section \ref{sec:conclusion}. Throughout the paper we use discrete time model $t \in \mathbb{N}_0 = \{0, 1, 2, \ldots \}$ and assume the random processes to be stationary, that is, the random processes depends only on the length of the interval $(s, t]$ ($\Delta = t-s$) , but not on $s$ or $t$ itself.

%Though the results discussed in this paper can be shown to be valid in a general setting, the stationarity of the stochastic processes is assumed to simplify the presentation of the paper. 

\section{Statistical Network Calculus}
\label{sec:SNC}
In this section, we give a brief overview of the statistical network calculus and our notion of statistical service envelope. Then, we derive performance bounds in using our notion of statistical service envelope.

The elegant theory of network calculus  \cite{boudec:2001,chang:2000} provides useful insights for the understanding of fundamental concepts of integrated and differentiated services, flow control, resource (bandwidth or buffer) dimensioning in data networks. The two main advantages of network calculus are ($i$) the relative ease in its ability to model different scheduling algorithms used at a network node, and ($ii$) the possibility to model a network of nodes as a single abstract node, which substantially reduces the complexity involved in network analysis. The mathematical theory of min-plus algebra forms the basis for the theory of network calculus. In the following we recall the two most commonly used min-plus operations in network calculus, namely, min-plus convolution and min-plus de-convolution operations \cite{boudec:2001}. 
\begin{definition}
Let $f(s,t)$ and $g(s,t)$ be two non-decreasing, real valued, bivariate functions defined at $t\ge s \ge 0$. Then the min-plus convolution ($\otimes$) and de-convolution ($\oslash$) operations are defined as follows:
\begin{eqnarray}
f \otimes g(s,t) &=& \inf_{s \le k \le t} \{f(s,k) + g(k,t)\}
\label{eq:minplus} \\
f \oslash g(s,t) &=& \sup_{0 \le k \le s} \{f(k,t) - g(k,s)\}
\label{eq:minplus_de}
\end{eqnarray}
\end{definition}

The statistical network calculus is the probabilistic version of network calculus and aims to profit from the statistical multiplexing in data networks. The fundamental difference between the statistical network calculus and its deterministic counterpart is that the performance bounds are expressed as probabilistic tail bounds, i.e., the derived bounds are violated with some probability. Arrival and departure processes are described using real-valued, bivariate functions $A(s,t)$ and $D(s,t)$, respectively, which represent the cumulative amount of data seen in the interval $(s,t]$ for any $0 \le s \le t$. We assume that there are no arrivals in the interval $(-\infty, 0]$ and $A(t) = A(0,t)$, $D(t) = D(0,t)$  for any $t \ge 0$. Since we assume the random process to be stationary, random process depends only on the length of the interval $(s, t]$, that is $\Delta = t-s$ , but not on $s$ or $t$ itself (therefore,  $A(s, t) = A(\Delta), D(s,t) = D(\Delta)$). For an arrival process $A$, a non-decreasing, real valued function is called statistical arrival envelope function or effective envelope ${\cal G}$ \cite{florin:2006} if the function satisfies the following condition, for any $t,s,\sigma \ge 0$:
\begin{equation}
P \{ A(s,t) > {\cal G}(t-s) + \sigma \} \le \varepsilon_g (\sigma)
\label{effenv}
\end{equation}
where $\varepsilon_g$ is called error function which is a non-negative, decreasing function of $\sigma$ bounding the violation probability. The sufficient condition for the derived performance bounds  to be finite is that the error function is required to satisfy the integrability condition \cite{florin:2006} given below:
\begin{equation}
\int^{\infty}_{0} \varepsilon(u)du<\infty
\label{int}
\end{equation}

Similarly, the service available to a flow is characterized using a non-decreasing, real valued function called statistical service envelope or effective service envelope ${\cal S}$ \cite{cruz:1996-1} such that for a given arrival process ($A$) and departure process ($D$), the service envelope satisfies the following condition for all $t,\sigma \ge 0$:
\begin{equation}
P \{ A\otimes{\cal S}(t) > D(t) + \sigma \} \le \varepsilon_s(\sigma)
\label{effsenv}
\end{equation}
where $\varepsilon_s$ is a decreasing error function bounding the violation probability. This error function is also required to satisfy the integrability condition from equation (\ref{int}) as the sufficient condition for the derived performance bounds to be finite. In this paper, we define a different notion of statistical service envelope (${\cal S}$) which is derived from the stochastic service process ($S$) characterizing the service offered at a network node. For a given arrival process ($A$) and departure process ($D$) at a network node, the stochastic service process ($S$) describing the service offered at the node satisfies the following condition for any fixed sample path and all $t \ge 0$:
\begin{equation}
 A\otimes S(t) \le D(t)
 \label{reffsenv}
\end{equation}
The key observation is that the stochastic service process ($S$) for any fixed sample path is a deterministic service envelope of the service offered to the given arrival process trajectory, and equation (\ref{reffsenv}) follows from the definition of deterministic service envelope \cite{boudec:2001}.
%Since the deterministic service envelope can be interpreted as the deterministic service process characterizing the worst case service offered at a network node, the definition of deterministic service envelope closely resemble the condition from equation ($\ref{reffsenv}$) for statistical service process. 
Any random process $S$ satisfying the above relationship (equation (\ref{reffsenv})) between arrival process and departure process for any fixed sample path is referred to as ``dynamic F-server'' \cite{chang:2000}. We now define a different notion of statistical service envelope (${\cal S}$) based on the stochastic service process ($S$) at the node.
\begin{definition}
Let $S$ be the stochastic service process characterizing the service offered at the node, then the statistical service envelope ${\cal S}$ for all $t,s, \sigma \ge 0$ can be defined as:
\begin{equation}
P \{ S(s,t) < {\cal S}(t-s) - \sigma \} \le \varepsilon_s (\sigma)
\label{nreffsenv} 
\end{equation}
where $\varepsilon_s$ is a decreasing error function bounding the violation probability and fulfilling the integrability condition from equation (\ref{int}).
\end{definition}
%Another variation of the definition of statistical service envelope from equation (\ref{nreffsenv}) with a positivity requirement for all $t,s, \sigma \ge 0$ is given by 
%\begin{equation}
%P \{ S(s,t) < [{\cal S}(t-s) - \sigma]^+ \} \le \varepsilon_s (\sigma)
%\label{nreffsenv11} 
%\end{equation}
%We use the notion $[X]^+ = \max(0,X)$ to denote the positive part of the real number $X$. It can be shown that the statistical service envelope from equation (\ref{nreffsenv11}) provides tighter performance bounds, but we prefer the statistical service envelope from equation (\ref{nreffsenv}) due to its inherent mathematical simplicity. 
The new definition of statistical service envelope does not imply equation (\ref{effsenv}). However, the sample path bound of the statistical service envelope from equation (\ref{nreffsenv}) is also a valid bound to its counterpart from equation (\ref{effsenv}), i.e., for all $t \ge 0$
\begin{eqnarray*}
P\left\{ A \otimes {\cal S}(t) > D(t) + \sigma \right\} &\le& P\left\{ A \otimes {\cal S}(t) > A\otimes S(t) + \sigma \right\} \\
&\le & P\left\{\sup_{0\le k \le t}\left\{ {\cal S}(t-k) - S(k,t) \right\} > \sigma \right\}\\
\mbox{and, } P\left\{ {\cal S}(t) - S(t) >  \sigma  \right\} & \le & P\left\{\sup_{0\le k \le t}\left\{ {\cal S}(t-k) - S(k,t) \right\} > \sigma \right\}
\end{eqnarray*}
The main advantage of the new definition of statistical service envelope from equation (\ref{nreffsenv}) is that its sample path bound is a lower bound to sample bound bound of statistical service envelope from equation (\ref{effsenv}), which is necessary information required about the service offered in a network to compute end-to-end performance measures, i.e.,
\begin{eqnarray*} 
\lefteqn{P\left\{\sup_{0\le k \le t}\left\{ {\cal S}(t-k) - S(k,t) \right\} > \sigma \right\}} \ \ \ \ \ \ \ \ \ \ \ \ \ \ \ \ \ \ \ \ \ \ \ \ \ \ \ \ \ \ \ \ \ \ \ \ \\
 & \le & P\left\{ \sup_{0\le k \le t}\left\{ A \otimes {\cal S}(t-k) -  D(k,t) \right\} > \sigma \right\}
\end{eqnarray*} 
The sample path bound of the statistical service envelope from equation (\ref{nreffsenv}) is given by the following lemma.
\begin{lemma}
\label{lemma:speffsenv}
Consider the service offered at a network node being described using a stochastic service process $S$ and let ${\cal S}$ be the statistical service envelope derived from stochastic service process $S$ satisfying equation (\ref{nreffsenv}) with an error function $\varepsilon_s$ satisfying integrability condition from equation (\ref{int}). Then for any $\delta >0$ and all $t,  \sigma \ge 0$
\begin{equation}
P\left\{\sup_{0\le k \le t}\left\{ {\cal S}(t-k) - S(k,t)  - \delta \cdot (t-k) \right\} > \sigma \right\} \le \sum_{u=0}^{\infty} \varepsilon_s(\sigma + \delta u)
\label{speffsenv} 
\end{equation}
\end{lemma}
The term $\delta$ is used in the above lemma to make the violation probability function of the envelope dependent on time, so that we can compute the sample path violation probability in terms of the given violation probability function of service envelope from equation (\ref{nreffsenv}). Intuitively, the term $\delta$ can be seen as rate correction factor that reduces the guaranteed service by a rate $\delta$.
\textbf{\textit{Proof:}} For a given $t, \sigma \ge 0$ and $\delta > 0$ we have
\begin{eqnarray*}
\lefteqn{P\left\{\sup_{0\le k \le t}\left\{ {\cal S}(t-k) - S(k,t) - \delta \cdot (t-k) \right\} > \sigma \right\}} \ \ \ \ \ \ \ \ \ \ \ \ \ \ \ \ \ \ \ \ \ \ \ \ \ \ \ \ \ \ \ \ \ \ \ \ \ \ \ \ \ \ \ \ \ \ \ \ \ \\
 &\le& \sum_{u=0}^{\infty} P\left\{ {\cal S}(u) - (\delta u + \sigma) > S(u) \right\}\\
&\le& \sum_{u=0}^{\infty} \varepsilon_s(\sigma + \delta u)
\end{eqnarray*}
The first inequality is due to the application of Boole's inequality and setting $u = t-k$. The second inequality is from the definition of statistical service envelope from equation (\ref{nreffsenv}). $\blacksquare$  

The new definition of statistical service envelope from equation (\ref{nreffsenv}) will be shown to be beneficial for the end-to-end network analysis in Section \ref{sec:scale} of this paper. From now on, unless specified otherwise, the term statistical service envelope refers to its definition from equation (\ref{nreffsenv}).

\begin{figure}
\centering
\includegraphics[scale=0.45]{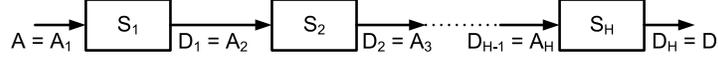}
\caption{ Network of H concatenated nodes}
\label{fig:tandemnet} % caption for the whole figure
\vspace{-5 mm}
\end{figure}

We now state our main results using the statistical service envelope derived from the stochastic service process at the network node. Consider a flow traversing through a network of $H$ nodes connected in series as shown in Fig. \ref{fig:tandemnet}. We assume that the service available for a flow at each hop ($h$) is characterized by a stochastic service process ($S_h$). The following theorem will provide a possibility to characterize the service offered by a network of nodes as shown in Fig. \ref{fig:tandemnet} in-terms of per-node statistical service envelopes.

\begin{theorem}
\label{theorem:snet}
Consider a flow traversing a network of $H$ nodes connected in series with each hop ($h=1, 2, \ldots, H$) offering a service characterized by its corresponding stochastic service process ($S_h=S_1, S_2, \ldots, S_H$). Then, the stochastic network service process $S_{net}$ for any fixed sample path is given by 
\begin{equation}
S_{net} = S_1 \otimes S_2 \otimes \cdots \otimes S_H
\label{netreffsenv} 
\end{equation}
and the corresponding statistical network service envelope ${\cal S}_{net}$ is given by 
\begin{equation}
{\cal S}_{net} = {\cal S}_{1} \otimes {\cal S}_{2} \otimes \cdots \otimes {\cal S}_{H}
\label{neteffsenv} 
\end{equation}
with a decreasing error function $\varepsilon_{s_{net}}$, for any $\delta > 0$, given by
\begin{equation}
\varepsilon_{s_{net}}(\sigma) = \inf_{\sigma_1 + \cdots + \sigma_H = \sigma} \left\{ \sum_{h=1}^{H} \sum_{u=0}^{\infty} \varepsilon_{s_{h}}(\sigma_h + \delta u)  \right\}
\label{neterror} 
\end{equation}
such that the statistical network service envelope ${\cal S}_{net}$ satisfies the following condition, for any $t, \sigma \ge 0$ and $\delta > 0$, 
\begin{equation}
P \left\{ \sup_{0 \le k \le t} \left\{ {\cal S}_{net}(t-k) - S_{net}(k,t) - \delta \cdot (t-k) \right\} >  \sigma \right\} \le \varepsilon_{s_{net}}(\sigma)
\end{equation}
\end{theorem}
%The term $\delta$ is similar to the rate correction factor introduced in \cite{florin:2006}. However, since we use a different definition of the statistical service envelope, the rate correction factor at each hop $h$ remains the constant $\delta$, in contrast to $(h-1)\delta$ at hop $h$ for the statistical service envelope defined using equation (\ref{effsenv})\cite{florin:2006}.\vspace{+0.3cm}\\
%the factor $\delta$ influences only the error function and has no effect on the service envelope of individual nodes in the network, in contrast to the non-random statistical service envelope defined using equation (\ref{effsenv})\cite{florin:2006}, where the service envelope at the $h$-th node is reduced by a rate $(h-1)\delta$. 
%\vspace{+0.3cm}\\
\textbf{\textit{Proof:}}
The proof of the theorem has two parts; The first is to prove the stochastic network service process and the second part is to prove the statistical network service envelope. Let $A=A_1$ be the arrival traffic at the node $1$ or ingress of the network and $D=D_{H}=A_{H+1}$ represent the departure traffic from the network of $H$ nodes connected in series as shown in Fig. \ref{fig:tandemnet}. The departure traffic $D_{h}$ from the node at hop $h$ becomes the arrival traffic $A_{h+1}$ to the downstream node at hop $h+1$, i.e., $A_{h+1}=D_h$ for all $h=1,\ldots, H$. 

In order to prove, for any sample path the stochastic network service process $S_{net} = S_1\otimes S_2 \otimes \ldots \otimes S_H$ represents the service offered by the network shown in Fig. \ref{fig:tandemnet}, one needs to show that the departure traffic $D$ from the network satisfies for any sample path the condition $ D \ge A \otimes S_{net}$. This can be shown in a straightforward fashion. From the property of stochastic service process characterizing the service offered at a node (equation (\ref{reffsenv})), the respective departure traffic satisfies for any sample path the condition $D_{h} \ge A_h \otimes S_h$ for $h = 1, 2, \ldots, H$, where $S_1, S_2, \ldots, S_H$ represents the stochastic network service process offered by  the respective $H$ nodes in the network. Applying the condition on departure traffic from each node iteratively for the departure traffic $D=D_H$ from the network, one gets for any sample path $D \ge A \otimes S_1 \otimes S_2 \otimes \ldots \otimes S_H = A \otimes S_{net}$. This proves our first claim.
For the given $t,\sigma \ge 0$ and $\delta > 0$, we have 
\begin{eqnarray*}
\lefteqn{P\left\{\sup_{0 \le k_1 \le t}\left\{{\cal S}_{net}(t-k_1) -  S_{net}(k_1,t)  - \delta \cdot (t-k_1)\right\} > \sigma\right\}}\\
&=& P\left\{ \sup_{0 \le k_1 \le t}\left\{ {\cal S}_{1}\otimes {\cal S}_{2}\otimes  \cdots \otimes{\cal S}_{H}(t-k_1) - S_1\otimes S_2 \otimes \cdots \otimes S_H(k_1,t) \right.\right. \\
&& \ \ \ \ \ \ \ \ \ \ \ \ \ \ \ \ \ \ \ \ \ \ \ \ \ \ \ \ \ \ \ \ \ \ \ \ \ \ \ \ \ \ \ \ \ \ \ \ \ \ \ \ \ \ \ \ \  \ \ \ \ \ \ \ \ \ \left. \left. -  \delta \cdot (t-k_1) \right\} > \sigma \right\} \\
&\le& P\left\{ \sup_{0 \le k_1 \le k_2 \le k_3 \cdots \le k_H \le t}\left\{{\cal S}_{1}(k_2-k_1) - S_1(k_1,k_2) - \delta \cdot (k_2-k_1)  \right.\right. \\
&& \ \ \ \ \ \ \ \ \ \ \ \ \ \ \ \ \ \ \ \ \ \ \ \ {\left.\left. + {\cal S}_{2}(k_3-k_2) - S_2(k_2,k_3) - \delta \cdot (k_3-k_2) + \cdots  \right.\right.} \\
&& \ \ \ \ \ \ \ \ \ \ \ \ \ \ \ \ \ \ \ \ \ \ \ \ \ \ \ \ \ \ \ {\left.\left. + {\cal S}_{H}(t-k_H) - S_H(k_H,t) - \delta \cdot (t-k_H) \right\} >  \sigma  \right\}}\\
&\le& P\left\{ \sup_{0 \le k_1 \le k_2 \le t}\left\{{\cal S}_{1}(k_2-k_1) - S_1(k_1,k_2) - \delta \cdot (k_2-k_1) \right\} + \right.\\
&& \ \ \ \ \ \ \ \ \ \ \ {\left. + \sup_{0 \le k_2 \le k_{3} \le t}\left\{{\cal S}_{2}(k_3-k_2) - S_2(k_2,k_3) - \delta \cdot (k_3-k_2) \right\} + \cdots  \right.}\\
&& \ \ \ \ {\left. + \sup_{0 \le k_H \le t}\left\{{\cal S}_{H}(t-k_H) - S_H(k_H,t) - \delta \cdot (t-k_H) \right\} >  \sigma_1 + \cdots + \sigma_H  \right\}}\\
&\le& P\left\{ \sup_{0 \le k_1 \le k_2 \le t}\left\{{\cal S}_{1}(k_2-k_1) - S_1(k_1,k_2) - \delta \cdot (k_2-k_1) \right\} > \sigma_1 \right\}  \\ 
&& \ \ + P\left\{ \sup_{0 \le k_2 \le k_3 \le t}\left\{{\cal S}_{2}(k_3-k_2) - S_2(k_2,k_3) - \delta \cdot (k_3-k_2) \right\} > \sigma_2 \right\} + \cdots \\
&& \ \ \ \ \ \ \ \ \ \ \ \ \ + P\left\{ \sup_{0 \le k_H \le t}\left\{{\cal S}_{H}(t - k_H) - S_H(k_H,t) - \delta \cdot (t-k_H) \right\} > \sigma_H \right\}\\
&\le& \inf_{\sigma_1 + \cdots + \sigma_H = \sigma} \left\{ \sum_{h=1}^{H} \sum_{k=0}^{\infty} \varepsilon_{s_{h}}(\sigma_h + \delta k)  \right\} = \varepsilon_{s_{net}}(\sigma) 
\end{eqnarray*} 
The inequality in the third step is due to the property of supremum operation, i.e., $ \sup_{0\le s \le t} \{X(s) + Y(s)\}$ $ \le $ $\sup_{0\le s \le t} \{X(s)\}$ $+$ $\sup_{0\le s \le t} \{Y(s)\}$ \cite{yuming:2006}. The final inequality is from the definition of sample path statistical service envelope (equation (\ref{speffsenv})) and the stationarity assumption of stochastic service processes. $\blacksquare$\\
If the error function $\varepsilon_{s_{h}}$ of the individual statistical service envelope ${\cal S}_h$ at hop $h$ for $h = 1, 2, \ldots, H$ satisfies the integrability condition from equation (\ref{int}), then the error function $\varepsilon_{s_{net}}$ of the statistical network service envelope will be finite (i.e., $ \varepsilon_{s_{net}}  < \infty$).

We next describe the probabilistic performance bounds on backlog, delay and output burstiness using the statistical arrival envelope (equation (\ref{effenv})) and network service envelope (Theorem \ref{theorem:snet}). In the following theorem, we use the notation ${\cal S}_{h,-\delta}(t) = {\cal S}_h(t) - \delta t$ and ${\cal G}_{\delta}(t) = {\cal G}(t) + \delta t$ to simplify the presentation. 
\begin{theorem}
\label{theorem:pbnr}
Let $A$ and $D$ be the arrival and departure traffic, respectively, from a network of $H$ nodes connected in series and ${\cal G}$ be the corresponding statistical arrival envelope with an error function $\varepsilon_g$. Assume $S_{net}$ is the stochastic network service process that characterizes the service offered by the network and ${\cal S}_{net}$ is the corresponding statistical network service envelope with an error function $\varepsilon_{s_{net}}$. Then we have the following bounds.
\begin{enumerate}
	\item Backlog bound : The probabilistic bound on the backlog in a network, for any $t, \sigma \ge 0$ and $\delta > 0$, is given by
				\begin{equation}
				P\left\{B(t) > {\cal G}_{\delta}\oslash{\cal S}_{net,-\delta}(0) + \sigma \right\} \le \varepsilon(\sigma)
				\label{backlognr} 
				\end{equation}	
	\item Delay bound : The probabilistic bound on the delay in a network, for any $t, \sigma \ge 0$ and $\delta > 0$, is given by
				\begin{equation}
				P\left\{W(t) > d(\sigma) \right\} \le \varepsilon(\sigma)
				\label{delaynr} 
				\end{equation}	
				where $d(\sigma) = \inf\{x: {\cal G}_{\delta}(t) + \sigma \le {\cal S}_{net, -\delta}(t+x)$ for all   $ t \ge 0\}$
	\item Output Burstiness : ${\cal G}_{\delta}\oslash{\cal S}_{net,-\delta}$ is a statistical arrival envelope of the departure traffic from the network, which satisfies the following condition for any $t,s, \sigma \ge 0$ and $\delta > 0$:
				\begin{equation}
				P\left\{D(s,t) > {\cal G}_{\delta}\oslash{\cal S}_{net,-\delta}(t-s) + \sigma \right\} \le \varepsilon(\sigma)
				\label{oeffenv} 
				\end{equation}
\end{enumerate}
where the error function $\varepsilon$ is given by
	\begin{eqnarray}
		\varepsilon(\sigma) &=& \inf_{\sigma_g + \sigma_{s_{net}} = \sigma} \left\{  \sum_{k=0}^{\infty} \varepsilon_g(\sigma_g + \delta k) + \varepsilon_{s_{net}}(\sigma_{s_{net}}) \right\}\\
		&=&  \inf_{\sigma_g + \sigma_{s_1} + \cdots + \sigma_{s_H} = \sigma} \left\{  \sum_{k=0}^{\infty} \varepsilon_g(\sigma_g + \delta k) + \sum_{h=1}^{H} \sum_{k=0}^{\infty} \varepsilon_{s_{h}}(\sigma_{s_h} + \delta k) \right\} \label{error} 
	\end{eqnarray}
\end{theorem}
The proof of the theorem relies on the sample path bound of statistical arrival envelope (equation (\ref{effenv})) for all $t, \sigma \ge 0$ and any $\delta > 0$ given by \cite{florin:2006}.
\begin{equation}
P\left\{\sup_{0\le k \le t}\left\{ A(k,t) - {\cal G}(t-k)  - \delta (t-k) \right\} > \sigma \right\} \le \sum_{u=0}^{\infty} \varepsilon_g(\sigma + \delta u)
\label{speffenv} 
\end{equation}
\textbf{\textit{Proof:}}
We now prove the probabilistic bound on backlog $B(t)$, for some $t \ge 0$. The backlog $B(t)$ at a network node is given as $A(t) - D(t)$. Therefore, for all $t, \sigma \ge 0$ and any $\delta > 0$, we have 
\begin{eqnarray*}
\lefteqn{P\left\{ B(t) > {\cal G}_{\delta}\oslash{\cal S}_{net,-\delta}(0) + \sigma \right\}}\\
&=&  P\left\{ A(t) - D(t) > {\cal G}_{\delta}\oslash{\cal S}_{net,-\delta}(0) + \sigma \right\}\\
&\le& P\left\{ A(t) - A\otimes S_{net}(t) - {\cal G}_{\delta}\oslash{\cal S}_{net,-\delta}(0) > \sigma \right\}\\
&\le& P\left\{ \sup_{0 \le k \le t} \left\{ A(t) - A(k) - {\cal G}_{\delta}(t-k) - S_{net}(k,t) + {\cal S}_{net,-\delta}(t-k) \right\} > \sigma \right\}\\
&\le&  P\left\{ \sup_{0 \le k \le t} \left\{ A(k,t) - {\cal G}(t-k) - {\delta} (t-k) \right\} \right.\\
&& \ \ \ \ \ \ \ \ \ \ \ \ \ \ \ \ \left. + \sup_{0 \le k \le t} \left\{{\cal S}_{net}(t-k) - S_{net}(k,t) - {\delta} (t-k) \right\} > \sigma_g + \sigma_{s_{net}} \right\}\\
&\le&  P\left\{ \sup_{0 \le k \le t} \left\{ A(t) - A(t-k) - {\cal G}(k) - {\delta} k \right\} > \sigma_g \right\} \\
&& \ \ \ \ \ \ \ \ \ \ \ \ \ \ \ \ \ \ \ \ \ \ \ \ \ \ \ \ \ \ \ \ +  P\left\{ \sup_{0 \le k \le t} \left\{{\cal S}_{net}(k) - S_{net}(k) - {\delta} k \right\} > \sigma_{s_{net}} \right\}\\
&\le& \inf_{\sigma_g + \sigma_{s_{net}} = \sigma} \left\{  \sum_{u=0}^{\infty} \varepsilon_g(\sigma_g + \delta u) + \varepsilon_{s_{net}}(\sigma_{s_{net}}) \right\}\\
&=&  \inf_{\sigma_g + \sigma_{s_1} + \cdots + \sigma_{s_H} = \sigma} \left\{  \sum_{u=0}^{\infty} \varepsilon_g(\sigma_g + \delta u) + \sum_{h=1}^{H} \sum_{u=0}^{\infty} \varepsilon_{s_{h}}(\sigma_{s_h} + \delta u) \right\}
\end{eqnarray*}
The third inequality is due to the property of supremum operation, i.e., $ \sup_{0\le s \le t} \{X(s) + Y(s)\}$ $ \le $ $\sup_{0\le s \le t} \{X(s)\}$ $+$ $\sup_{0\le s \le t} \{Y(s)\}$ \cite{yuming:2006}. The final inequality is from the definition of statistical network service envelope (Theorem \ref{theorem:snet}) and sample path statistical arrival envelope (equation \ref{speffenv}). The proofs of the probabilistic bounds on delay and departure process are the immediate variations of the proof presented above and  are omitted. $\blacksquare$  \\
%It should be noted that the error function $\varepsilon$ may not satisfy the integrability condition from equation (\ref{int}). However, if the error function $\varepsilon_{s_{h}}$ of the individual statistical service envelope ${\cal S}_h$ at hop $h$ for $h = 1, 2, \ldots, H$  and  the error function $\varepsilon_{g}$ of statistical arrival envelope ${\cal G}$ satisfy the integrability condition from equation (\ref{int}), then the error function $\varepsilon$ of the probabilistic performance bounds will be finite (i.e., $ \varepsilon  < \infty$). 
The optimal value of $\delta$ is chosen to minimize the violation probability of the performance bounds. The probabilistic performance bounds from Theorem \ref{theorem:pbnr} can be further improved if the arrival traffic process $A$ and the stochastic service process $S_h$ at each hop $h$ for $h = 1, 2, \ldots, H$ are statistically independent of one another.  In the following theorem, we use the notation for conventional convolution $\varepsilon_g * \varepsilon_s(\sigma)= \int_{0}^{\sigma}{\varepsilon_g(\sigma-u)d\varepsilon_s(u)}$ (as error functions are non-negative functions) to simplify the presentation. 

\begin{theorem}
\label{theorem:pbnr1}
Let $A$ be the arrival traffic independent of the service offered at the network of $H$ nodes connected in series and ${\cal G}$ is the corresponding statistical arrival envelope with an error function $\varepsilon_g$. Assume $S_h$ is the stochastic service process characterizing the service offered at the hop $h$ for $h = 1, 2, \ldots, H$ and the services offered at each hop are statistically independent of one another. Let ${\cal S}_{h}$ be the corresponding statistical service envelope with an error function $\varepsilon_{s_{h}}$ at hop $h$ for $h = 1, 2, \ldots, H$ and ${\cal S}_{net}$ be the statistical network service envelope. Then we have the following bounds.
\begin{enumerate}
	\item Backlog bound : The probabilistic bound on the backlog in a network, for any $t, \sigma \ge 0$ and $\delta > 0$, is given by
				\begin{equation}
				P\left\{B(t) > {\cal G}_{\delta}\oslash{\cal S}_{net,-\delta}(0) + \sigma \right\} \le 1-(\tilde{\varepsilon}_g * \tilde{\varepsilon}_{s_1}* \tilde{\varepsilon}_{s_2}* \cdots * \tilde{\varepsilon}_{s_H})(\sigma)
				\label{backlognr1} 
				\end{equation}	
	\item Delay bound : The probabilistic bound on the delay in a network, for any $t, \sigma \ge 0$ and $\delta > 0$, is given by
				\begin{equation}
				P\left\{W(t) > d(\sigma) \right\} \le 1-(\tilde{\varepsilon}_g * \tilde{\varepsilon}_{s_1}* \tilde{\varepsilon}_{s_2}* \cdots * \tilde{\varepsilon}_{s_H})(\sigma)
				\label{delaynr1} 
				\end{equation}	
				where $d(\sigma) = \inf\{x: {\cal G}_{\delta}(t) + \sigma \le {\cal S}_{net, -\delta}(t+x)$ for all   $ t \ge 0\}$
	\item Output Burstiness : ${\cal G}_{\delta}\oslash{\cal S}_{net,-\delta}$ is a statistical arrival envelope of the departure traffic $D$ from the network, which satisfies the following condition for any $t,s, \sigma \ge 0$ and $\delta > 0$:
				\begin{equation}
				P\left\{D(s,t) > {\cal G}_{\delta}\oslash{\cal S}_{net,-\delta}(t-s) + \sigma \right\} \le 1-(\tilde{\varepsilon}_g * \tilde{\varepsilon}_{s_1}* \tilde{\varepsilon}_{s_2}* \cdots * \tilde{\varepsilon}_{s_H})(\sigma)
				\label{oeffenv1} 
				\end{equation}
\end{enumerate}
where $\tilde{\varepsilon}_g(\sigma) = 1 - \sum_{k=0}^{\infty} \varepsilon_g(\sigma + \delta k)$  and $\tilde{\varepsilon}_{s_h}(\sigma) = 1 - \sum_{k=0}^{\infty} \varepsilon_{s_h}(\sigma + \delta k)$ for $h = 1, 2, \ldots, H$.
\end{theorem}
The proof of the theorem relies on the Lemma 4.1 from \cite{yuming:2006}, which states that for any two non-negative independent random variables $F$ and $G$ with $P(F > \sigma) \le f(\sigma)$ and $P(G > \sigma) \le g(\sigma)$ where $f(\sigma)$ and $g(\sigma)$ are non-negative, decreasing function for any $\sigma \ge 0$, then 
\begin{equation}
P\left\{F + G > \sigma \right\} \le 1-(\tilde{f}*\tilde{g})(\sigma)
\label{speffenv1} 
\end{equation}
where $\tilde{f}(\sigma) = 1 - f(\sigma)$ and $\tilde{g}(\sigma) = 1 - g(\sigma)$. \vspace{+0.3cm}\\
\textbf{\textit{Proof:}}
We now prove the probabilistic bound on backlog $B(t)$, for some $t \ge 0$. The backlog $B(t)$ at a network node is given as $A(t) - D(t)$. Therefore, for all $t, \sigma \ge 0$ and any $\delta > 0$, we have 
\begin{eqnarray*}
\lefteqn{P\left\{ B(t) > {\cal G}_{\delta}\oslash{\cal S}_{net,-\delta}(0) + \sigma \right\}}\\
&=&  P\left\{ A(t) - D(t) > {\cal G}_{\delta}\oslash{\cal S}_{net,-\delta}(0) + \sigma \right\}\\
&\le& P\left\{ A(t) - A\otimes S_{net}(t) - {\cal G}_{\delta}\oslash{\cal S}_{net,-\delta}(0) > \sigma \right\}\\
&\le&  P\left\{ \sup_{0 \le k \le t} \left\{ A(k,t) - {\cal G}(t-k) - {\delta} (t-k)  \right. \right. \\
&& \ \ \ \ \ \ \ \ \ \ \ \ \ \ \ \ \ \ \ \ \ \ \ \ \ \ \ \ \ \ \ \ \ \ \left. \left.+  {\cal S}_{net}(t-k) - S_{net}(k,t) - {\delta} (t-k) \right\} > \sigma \right\}\\
&=& P\left\{ \sup_{0 \le k \le t} \left\{ A(k,t) - {\cal G}(t-k) - {\delta} (t-k)  -  \delta (t-k) + \right. \right.\\
&& \ \ \ \ \ \ \ \ \ \ \left. \left.{\cal S}_{1}\otimes {\cal S}_{2}\otimes  \cdots \otimes{\cal S}_{H}(t-k) - S_1\otimes S_2 \otimes \cdots \otimes S_H(k,t) \right\} > \sigma \right\}\\
&\le& P\left\{ \sup_{0 \le k \le t} \left\{ A(k,t) - {\cal G}(t-k) - {\delta} (t-k) \right\} \right.\\
&& \ \ \ \left. + \sup_{0 \le k \le k_2 \le t}\left\{{\cal S}_{1}(k_2-k) - S_1(k,k_2) - \delta (k_2-k) \right\}  \right.\\
&& \ \ \ \left. +  \sup_{0 \le k_2 \le k_3 \le t}\left\{{\cal S}_{2}(k_3-k_2) - S_2(k_2,k_3) - \delta (k_3-k_2) \right\} \right. + \cdots    \\
&& \ \ \ \ \ \ \ \ \ \ \ \ \ \ \ \ \ \ \ \ \ \left. + \sup_{0 \le k_H \le t}\left\{{\cal S}_{H}(t-k_H) - S_H(k_H,t) - \delta (t-k_H) \right\} > \sigma \right\}\\
&\le& 1-(\tilde{\varepsilon}_g * \tilde{\varepsilon}_{s_1}* \tilde{\varepsilon}_{s_2}* \cdots * \tilde{\varepsilon}_{s_H})(\sigma)
\end{eqnarray*}
The third inequality is due to the property of supremum operation, i.e., $ \sup_{0\le s \le t} \{X(s) + Y(s)\}$ $ \le $ $\sup_{0\le s \le t} \{X(s)\}$ $+$ $\sup_{0\le s \le t} \{Y(s)\}$ \cite{yuming:2006}. The final inequality follows from equations  (\ref{speffsenv}), (\ref{speffenv}) and (\ref{speffenv1}). The proofs of the probabilistic bounds on delay and departure process are the immediate variations of the proof presented above and  are omitted. $\blacksquare$  \\ 
The performance bounds from Theorem \ref{theorem:pbnr} and Theorem \ref{theorem:pbnr1} are derived using the statistical envelopes and require an additional rate correction factor $\delta$ whose value is chosen to minimize the violation probability of the performance bounds. 

Before we proceed to analyzing the scaling properties of the performance bounds derived in Theorem \ref{theorem:pbnr}, we need an important result on leftover statistical service envelope using a generalized scheduling model. The leftover service envelope is a generic envelope modeling the service available to a flow of interest which is left unused by its neighboring flows sharing the resources at a node. The concept of leftover service envelope was first introduced in deterministic setting \cite{boudec:2001,chang:2000} and then extended to stochastic domain in \cite{florin:2006,li:2007,fidler:2006}. It should be noted that leftover service envelope accounts for a pessimistic estimate of the offered service at a node for a flow of interest, as it characterizes offered service as the worst case service available to a low priority flow in a queue with static priority scheduling. The following theorem describes the leftover statistical service envelope derived with statistical service envelope from equation (\ref{nreffsenv}).

\begin{figure}
\centering
\includegraphics[scale=0.25]{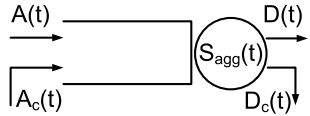}
\caption{ Network node with through and cross flows}
\label{fig:queue} % caption for the whole figure
\vspace{-5 mm}
\end{figure}

\begin{theorem}
\label{theorem:los}
Let ${\cal S}_{agg}$ be statistical service envelope with an error function $\varepsilon_{s_{agg}}$ and $S_{agg}$ be a stochastic service process characterizing the aggregate service offered at a queue with the flow of interest $A$ and the neighboring flow $A_c$ characterized using statistical arrival envelopes ${\cal G}$ and ${\cal G}_c$ with error functions $\varepsilon_{g}(u)$ and $\varepsilon_{g_c}(u)$, respectively. Let $D$ and $D_c$ be the departure flows from the queue for the corresponding flow of interest $A$ and the neighboring flow $A_c$, respectively.  Assuming that the stability condition at the queue is satisfied, i.e., $E\left[S_{agg}\right] \ge E\left[A\right] + E\left[A_c\right]$, then the leftover stochastic service process $S$, for any sample path and for all $t \ge 0$, is given by 
\begin{equation}
S(t) = S_{agg}(t) - A_c(t)
\label{rlos} 
\end{equation}
and the leftover statistical service envelope ${\cal S}$  with an error function $\varepsilon_{s}$, for all $t \ge 0$, is given by 
\begin{equation}
{\cal S}(t) = {\cal S}_{agg}(t) - {\cal G}_{c}(t)
\label{nrlos} 
\end{equation}
where $\varepsilon_{s}$ is given by
\begin{equation}
\varepsilon_{s}(\sigma) = \inf_{\sigma_{g_c} + \sigma_{s_{agg}} = \sigma} \left\{ \varepsilon_{g_c}(\sigma_{g_c}) + \varepsilon_{s_{agg}}(\sigma_{s_{agg}})  \right\}
\label{loserror} 
\end{equation}
\end{theorem}

\textbf{\textit{Proof:}}
From the property of stochastic service process (equation (\ref{reffsenv})) of the service offered at a queue shown in the Fig. \ref{fig:queue}, we have, for any sample path and for all $t \ge 0$

\begin{eqnarray*}
D(t) + D_c(t) &\ge& (A + A_c)\otimes S_{agg}(t)\\
D(t) &\ge& \inf_{0 \le k \le t} \left\{ A(k) + A_c(k) + S_{agg}(k,t)\right\} - D_c(t) \\
&\ge& \inf_{0 \le k \le t} \left\{ A(k) - A_c(k,t) + S_{agg}(k,t)\right\}  \\ 
&=& A \otimes (S_{agg} - A_c)(t) \\
&=& A \otimes S(t)
\end{eqnarray*}
The inequality in the third step is due to the fact that for any sample path the departure traffic is always bounded by the arrival traffic, i.e., $D_c(t) \le A_c(t)$. This proves our claim about the leftover stochastic service process.\\
For a given $t \ge 0$, we have 
\begin{eqnarray*}
\lefteqn{P\left\{ {\cal S}(t) - S(t) > \sigma \right\}}\\
&=& P\left\{ {\cal S}(t) - S_{agg}(t) + A_c(t) > \sigma \right\}\\
&=& P\left\{ A_c(t) + {\cal S}(t) - {\cal S}_{agg}(t) + {\cal S}_{agg}(t) - S_{agg}(t)  > \sigma \right\}\\
&\le& P\left\{ A_c(t) - {\cal G}_c(t) > \sigma_{g_c} \right\} + P\left\{ {\cal S}_{agg}(t) - S_{agg}(t)  > \sigma_{agg} \right\}\\
&\le& \inf_{\sigma_{g_c} + \sigma_{s_{agg}} = \sigma} \left\{ \varepsilon_{g_c}(\sigma_{g_c}) + \varepsilon_{s_{agg}}(\sigma_{s_{agg}})  \right\}\\
&=& \varepsilon_{s}(\sigma)
\end{eqnarray*}
This proves our claim about the leftover non-random statistical service envelope. $\blacksquare$

For a work conserving queue shown in Fig. \ref{fig:queue} which is served at a constant rate $C$, the aggregate stochastic service process and statistical service envelope of the service offered at the queue is $C$, i.e., $S_{agg} = {\cal S}_{agg} = C$ with error function $\varepsilon_{s_{agg}}=0$. From Theorem \ref{theorem:los}, for all $t \ge 0$, we get the leftover stochastic service process for any sample path is $S(t) = Ct - A_c(t)$ and the leftover statistical service envelope is ${\cal S}(t) = Ct - {\cal G}_c(t)$ with the error function $\varepsilon_s = \varepsilon_{g_c}$. 
%It should be noted that there has been also similar performance bounds derived using statistical network calculus for different characterization of arrival and service envelopes in the literature \cite{yuming:2006,florin:2005,boorstyn:2000-1,boorstyn:2000-2,burchard:2006,li:2007}.  we re-state the performance bounds using our definition of non-random statistical network service envelope.
%Depending on the ordering relationship, a random process can be defined as an envelope function in two ways \cite{florin:phd}. First method is using the notion of stochastic ordering \cite{stoyan:1983}, where we say the random variable $X$ is stochastically smaller than random variable $Y$ if $P\{X > z\} \le P\{Y > z\}$. The second method is using the notion of almost surely ordering \cite{florin:phd}, where we say the random variable $X$ is almost surely smaller than random variable $Y$ if $P\{X > Y\} = 0$. The almost surely ordering implies stochastic ordering but the reverse may not be true.
\section{Scaling of End-to-End Performance Bounds}
\label{sec:scale}

\begin{figure}
\centering
\includegraphics[scale=0.45]{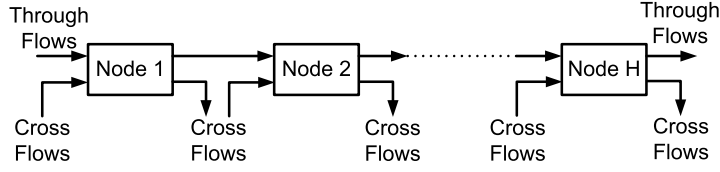}
\caption{ Network of H concatenated nodes with cross traffic}
\label{fig:tandemnet1} % caption for the whole figure
\vspace{-5 mm}
\end{figure}

In this section, we analyze the scaling properties of the performance bounds derived in Theorem \ref{theorem:pbnr}. For this purpose, as in \cite{fidler:2006}, we use the ($\sigma(\theta),\rho(\theta)$) - traffic model from \cite{chang:2000} in a network of $H$ nodes connected in series with cross traffic shown in Fig.\ref{fig:tandemnet1}. The flow of interest is the one which traverses through the network of $H$ nodes connected in series and is termed through flow $A$. The flow which transits the network at each hop is termed cross flow $A_c$. The network node at each hop has a work conserving scheduler which operates at a constant rate $C$. The goal is to determine the end-to-end performance (delay and backlog) bounds for the through flow in presence of the cross flow at each hop and identify its order of scaling. An arrival traffic $A$ with the effective bandwidth $\alpha$ is said to be a ($\sigma(\theta),\rho(\theta)$) - constrained arrival traffic if for any $t \ge 0$ and $\theta >0$ it satisfies the condition 
\begin{equation}
t \alpha(\theta, t) \le \rho(\theta) t + \sigma(\theta)
\label{sre}
\end{equation}
where $\rho(\theta)$ is a bound on the time independent version of effective bandwidth \cite{kelly:1996} $\left(\mbox{i.e., }\rho(\theta) \ge \lim_{t \rightarrow \infty} \alpha(\theta,t)\right)$. For $t \ge 0$, ${\cal G}(t) = \rho(\theta) t + \sigma(\theta)$ can be used as the statistical arrival envelope ${\cal G}$ of the arrival traffic $A$ with the error function $\varepsilon_g(x) =  e^{-\theta x}$, i.e., for all $t \ge s \ge 0$ and  any $\theta >0$ the following condition holds from Chernoff's bound and equation (\ref{sre}):
\begin{equation}
P\{A(s,t) > \rho(\theta) (t-s)  + \sigma(\theta) + \gamma\} \le e^{\theta (t-s) \alpha (\theta, t-s) - \theta \rho(\theta) (t-s) - \theta \sigma(\theta) - \theta \gamma} \le   e^{-\theta \gamma}
\end{equation}
Let the through flow $A$ with effective bandwidth $\alpha$ at the ingress of the network and the cross flow $A_c$ with effective bandwidth $\alpha_c$ at each hop be the ($\sigma(\theta),\rho(\theta)$) and ($\sigma(\theta),\rho_c(\theta)$) - arrival traffic with statistical arrival envelopes ${\cal G}(t) = \rho(\theta) t + \sigma(\theta)$ and ${\cal G}_c(t) = \rho_c(\theta) t + \sigma(\theta)$, respectively. For all $t \ge 0$ and any $\theta > 0$, the condition $C \ge \rho(\theta) + \rho_c(\theta)$ must be satisfied for stability. The stochastic arrival processes $A$ and $A_c$ describe the through flow and cross flow, respectively. The service available to the through flow at each hop can be characterized using leftover statistical service envelopes from Theorem \ref{theorem:los}. For all $t \ge 0$, the leftover stochastic service process $S_h$ for $h=1, \ldots, H$ is given as $S_h(t) = Ct- A_c(t)$, the leftover statistical service envelope ${\cal S}_h$ for $h=1, \ldots, H$ is given as ${\cal S}_h(t) = Ct- {\cal G}_c(t)$ with the error function $\varepsilon_{s_h}(\gamma) = \varepsilon_{g_c}(\gamma) = e^{-\theta \gamma}$. The statistical service envelope ${\cal S}_{net}(t)$ from Theorem \ref{theorem:snet}, for all $t \ge 0$, is given as ${\cal S}_{net}(t) = (C - \rho_c(\theta))t - H\sigma(\theta)$ with error function $\varepsilon_{s_{net}}(\gamma) = \sum_{h=1}^{H} \sum_{k=0}^{t} \varepsilon_{s_{h}}(\gamma+ \delta k)$. Throughout this section, we will evaluate the larger interval [$0,\infty$] instead of [$0,t$] to simplify the derivation of conservative, closed-form performance bounds. 

We first derive the end-to-end backlog $B(t)$ bound, for all $t \ge 0$, using the statistical envelopes. For any $\theta > 0$ and  $\frac{C-\rho(\theta)-\rho_c(\theta)}{2} \ge \delta \ge 0$, we get ${\cal G}_{\delta}\oslash{\cal S}_{net,-\delta}(0)$ = $(H+1) \sigma(\theta)$. The probabilistic backlog $B(t)$ bound from Theorem \ref{theorem:pbnr}, for all $t \ge 0$, is given as 
\begin{eqnarray}
\lefteqn{P\left\{B(t) > (H+1)\sigma(\theta) + \gamma \right\}} \nonumber \\
	&\le& \inf_{\gamma_g + \gamma_{s_1} + \cdots + \gamma_{s_H} = \gamma}   \sum_{k=0}^{\infty} \varepsilon_g(\gamma_g + \delta k) + \sum_{h=1}^{H} \sum_{k=0}^{\infty} \varepsilon_{s_{h}}(\gamma_{s_h} + \delta k) \nonumber \\
	&=& \inf_{\gamma_g + \gamma_{s_1} + \cdots + \gamma_{s_H} = \gamma}  \sum_{k=0}^{\infty}  e^{-\theta(\gamma_g + \delta k)} + \sum_{h=1}^{H} \sum_{k=0}^{\infty}  e^{-\theta(\gamma_{s_h} + \delta k)} \nonumber \\
	&=& \inf_{\gamma_g + \gamma_{s_1} + \cdots + \gamma_{s_H} = \gamma} \frac{1}{1-e^{-\theta \delta}} e^{-\theta \gamma_g} + \sum_{h=1}^{H} \frac{1}{1-e^{-\theta \delta}} e^{-\theta \gamma_{s_h}} \nonumber \\
	&=& \frac{(H+1)}{1-e^{-\theta \delta}} e^{-\frac{\theta \gamma}{H+1}} \label{e2eb1}
\end{eqnarray}
The final step is due to the convexity of $e^{-x}$. Usually we determine a backlog bound so that $P\{ B(t) > (H+1)\sigma(\theta) + \gamma \} \le \varepsilon$, where $\varepsilon$ is the given violation probability. Setting the right-hand side of equation (\ref{e2eb1}) to $\varepsilon$, using the optimal value of $\delta=\frac{C-\rho(\theta)-\rho_c(\theta)}{2}$  and solving for $\gamma$ gives, for any $\theta > 0 $
\begin{equation}
\gamma =  \frac{H+1}{\theta}\log{\frac{ (H+1)}{\varepsilon \left( 1-e^{-\frac{\theta(C-\rho(\theta)-\rho_c(\theta))}{2}} \right)}}  \label{e2eb21}
\end{equation}
Therefore the backlog $x = (H+1) \sigma(\theta) + \gamma$ can be explicitly bounded as follows:
\begin{equation}
x =  \inf_{\theta > 0} \frac{H+1}{\theta}\log{\frac{(H+1)}{\varepsilon \left( 1-e^{-\frac{\theta(C-\rho(\theta)-\rho_c(\theta))}{2}} \right)}} + (H+1) \sigma(\theta) \label{e2eb2}
\end{equation}
It is apparent from equation (\ref{e2eb2}) that the end-to-end backlog measure computed using Theorem \ref{theorem:pbnr} is bounded by ${\cal O}(H \log{H})$.

The same technique can be used to derive the end-to-end delay bound using Theorem \ref{theorem:pbnr}. For any $\theta > 0$ and $\frac{C-\rho(\theta)-\rho_c(\theta)}{2} \ge \delta \ge 0$, $d(\gamma)$ from equation (\ref{delaynr}) becomes $\frac{\gamma + (H+1)\sigma(\theta)}{(C-\rho_c(\theta) - \delta)}$, then the probabilistic bound on delay $W(t)$ from Theorem \ref{theorem:pbnr}, for all $t \ge 0$, is given as
\begin{equation}
P\left\{W(t) > d(\gamma) \right\} \le \frac{(H+1)}{1-e^{-\theta \delta}} e^{-\frac{\theta (C - \rho_c(\theta) - \delta)d(\gamma)}{H+1} + \theta \sigma(\theta) } \label{e2ed1}
\end{equation}
Usually we determine a delay bound so that $P\{ W(t) > d\} \le \varepsilon$, where $\varepsilon$ is the given violation probability and $d = d(\gamma)$. Setting the right-hand side of equation (\ref{e2ed1}) to $\varepsilon$, using the optimal value of $\delta$ as $\frac{C-\rho(\theta)-\rho_c(\theta)}{2}$  and solving for $d(\gamma)$ gives
\begin{equation}
d = \inf_{\theta > 0} \frac{2(H+1)}{\theta(C+\rho(\theta)-\rho_c(\theta))}\log{\frac{(H+1)}{\varepsilon \left( 1-e^{-\frac{\theta(C-\rho_c(\theta)-\rho(\theta))}{2}} \right)}}  + \frac{2(H+1) \sigma(\theta)}{(C+\rho(\theta)-\rho_c(\theta))} \label{e2ed2}
\end{equation}
It is apparent from equation (\ref{e2ed2}) that the end-to-end delay measure computed using Theorem \ref{theorem:pbnr} is bounded by ${\cal O}(H \log{H})$.

\section{Numerical Example}
\label{sec:num}
The goal of this section is to illustrate the benefits of using the new definition of statistical service envelope from equation (\ref{nreffsenv}) over its counterpart from equation (\ref{effsenv}) on the efficiency and scalability of the computed performance measures using a numerical example. For the numerical experiment we consider a network of $H$ concatenated nodes as shown in Fig. \ref{fig:tandemnet1}. The queue at each hop $h$ is served at a constant deterministic service rate $C$. We use the Markov modulated on-off (MMOO) process to describe the arrivals of $N$ independent through flows at the ingress of the network and the arrivals of $M$ independent cross flows at each hop $h$ inside the network. Markov modulated on-off process is a typical example of ($\sigma(\theta),\rho(\theta)$) - constrained traffic model with parameters ($0,\alpha(\theta)$) and is commonly used to model the voice \cite{voip_itu:1993} and video traffic \cite{maglaris:1988} in the Internet. Markov modulated on-off process can be in ``On" state or ``Off" state for a random time interval which is negative exponentially distributed with average $E[T_{on}]$ and $E[T_{off}]$, respectively. In ``On" state, arrival traffic transmits data at a constant rate $P$ and no data is transmitted in ``Off" state. The effective bandwidth of Markov modulated on-off process has an interesting property that $\alpha(\theta,t) \le  \alpha(\theta)$ and for any $\theta > 0$ is given by 
\begin{equation}
\alpha(\theta) = \frac{1}{2\theta}\left(P\theta-r_{{10}}-r_{{01}}+\sqrt {\left(P\theta - r_{{10}} + r_{{01}} \right)^{2} + 4r_{{10}}r_{{01}} }\right)
\end{equation}
where $r_{10} = \frac{1}{E[T_{on}]}$ and $r_{01} = \frac{1}{E[T_{off}]}$. 
In the example, we determine the numerical end-to-end delay bound for $N$ through flows in the network  with a violation probability $\varepsilon = 10^{-9}$. The capacity of the server $C$ at each hop is set to $100Mbps$. We use two types of Markov modulated on-off traffic model as in \cite{florin:2006}: $(i)$ ``high burstiness" variant has the parameters $E[T_{on}]=10ms$ and $E[T_{off}]=90ms$ and $(ii)$ ``low burstiness" variant has the parameters  $E[T_{on}]=1ms$ and $E[T_{off}]=9ms$. Both variants of the on-off traffic produce data at an average rate $m = 0.15Mbps$ and emit data at a peak rate $P = 1.5Mbps$ during the "`On"' state. The stochastic service available for $N$ through flows is determined using the generalized scheduling model (Theorem \ref{theorem:los}). 
\begin{figure}[ht]
\centering
\subfigure[with ``high burstiness" traffic]{
\includegraphics[angle=-90,scale=0.22]{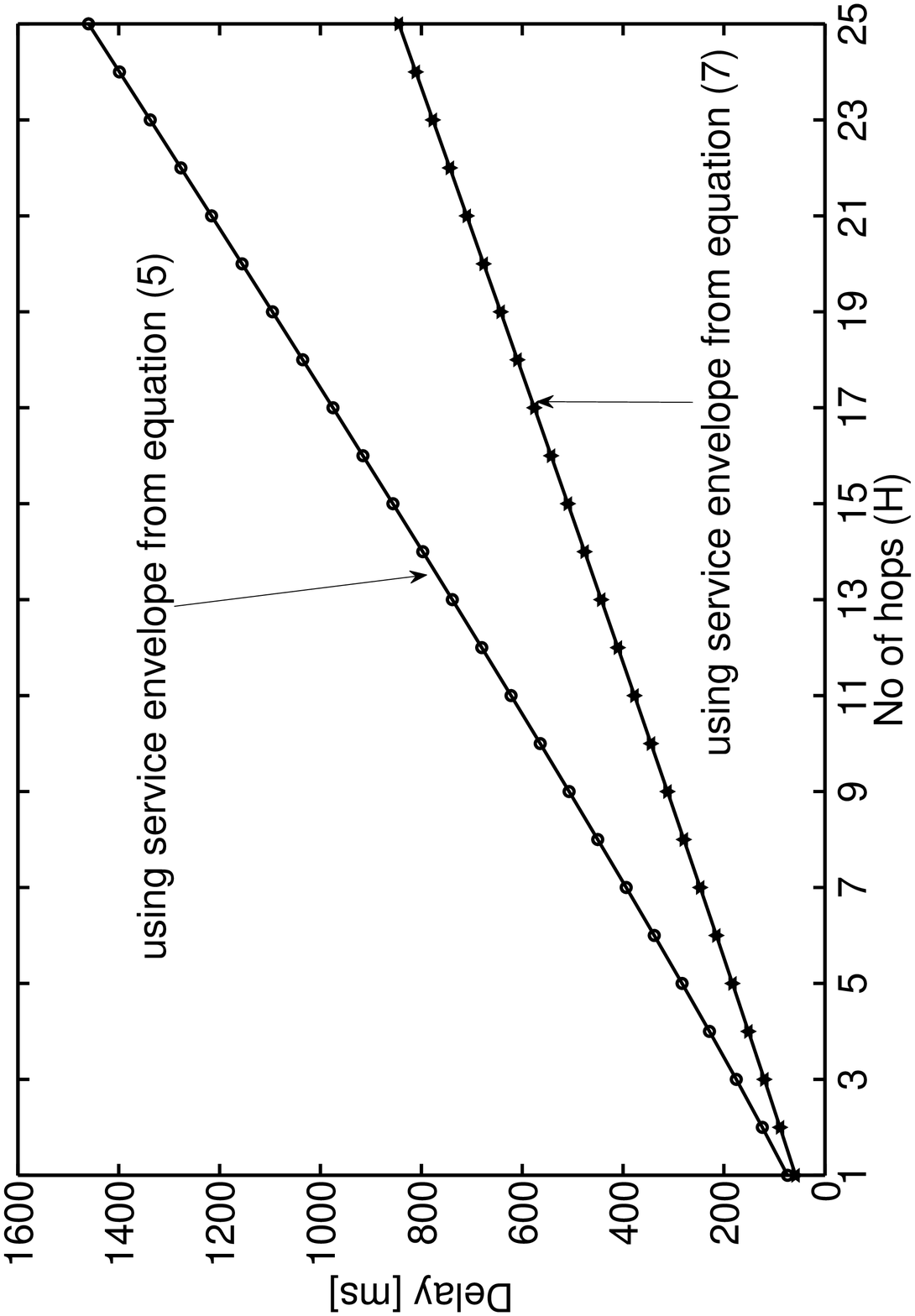}
\label{fig:plot1}
}
\subfigure[with ``low burstiness" traffic]{
\includegraphics[angle=-90,scale=0.22]{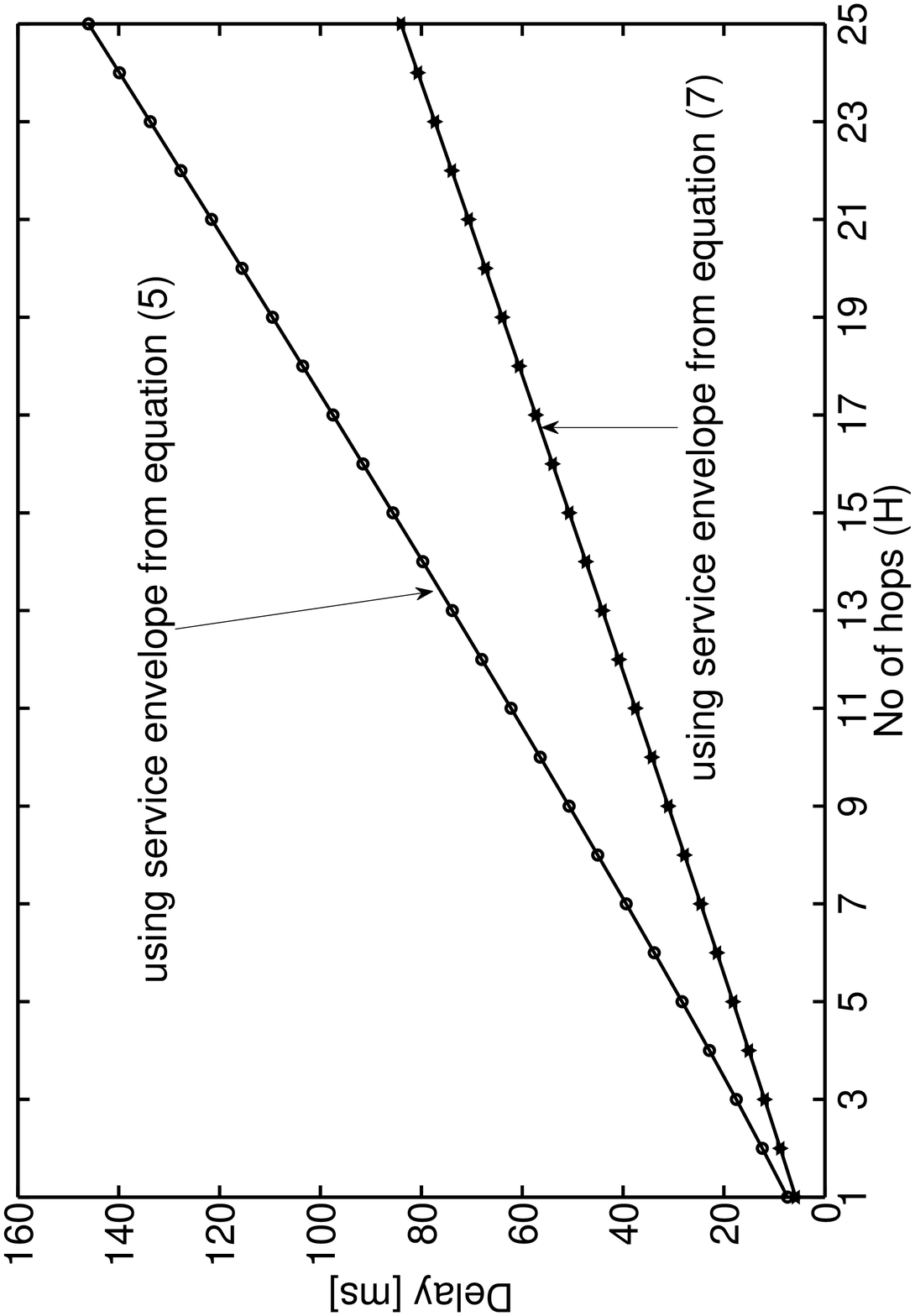}
\label{fig:plot2}
}
\label{fig:scaleplot}
\caption{End-to-end delay bound with a violation probability $\varepsilon = 10^{-9}$ for increasing number of hops $H$ with $N = 134$ through MMOO flows and $M=333$ cross MMOO flows at each hop}
\end{figure}

We compare the end-to-end delay bounds for through flows determined using the statistical service envelope from equation(\ref{nreffsenv}) and equation(\ref{effsenv}). For a validation of the latter we reproduced the results presented in \cite{florin:2006} using the statistical service envelope from equation(\ref{effsenv}). Fig. \ref{fig:scaleplot} shows the probabilistic end-to-end delay bounds with a violation probability $(\varepsilon)$ of $10^{-9}$ as a function of increasing number of hops $H$. At each hop, $M=333$ cross flows are multiplexed with $N = 134$ through flows. The plot illustrates the ${\cal O}(H \log{H})$ bounds of end-to-end delays with statistical network service envelope from Theorem \ref{theorem:snet} and validates that the end-to-end delay bounds determined using the statistical service envelope from equation (\ref{nreffsenv}) provide tighter bounds than the ones computed using the statistical service envelope from equation (\ref{effsenv}). 

\begin{figure}[ht]
\centering
\subfigure[with ``high burstiness" traffic]{
\includegraphics[angle=-90,scale=0.22]{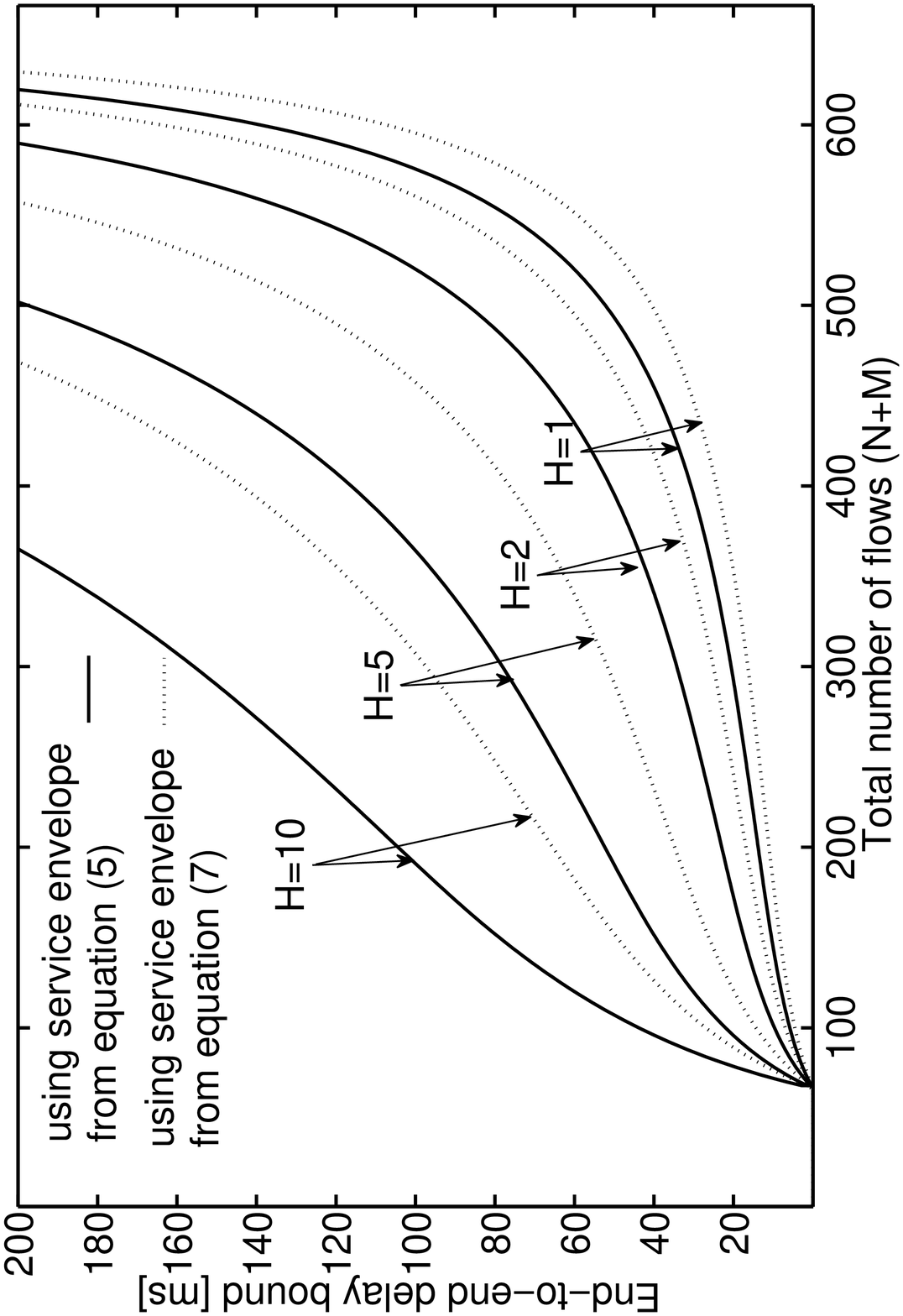}
\label{fig:plot3}
}
\subfigure[with ``low burstiness" traffic]{
\includegraphics[angle=-90,scale=0.22]{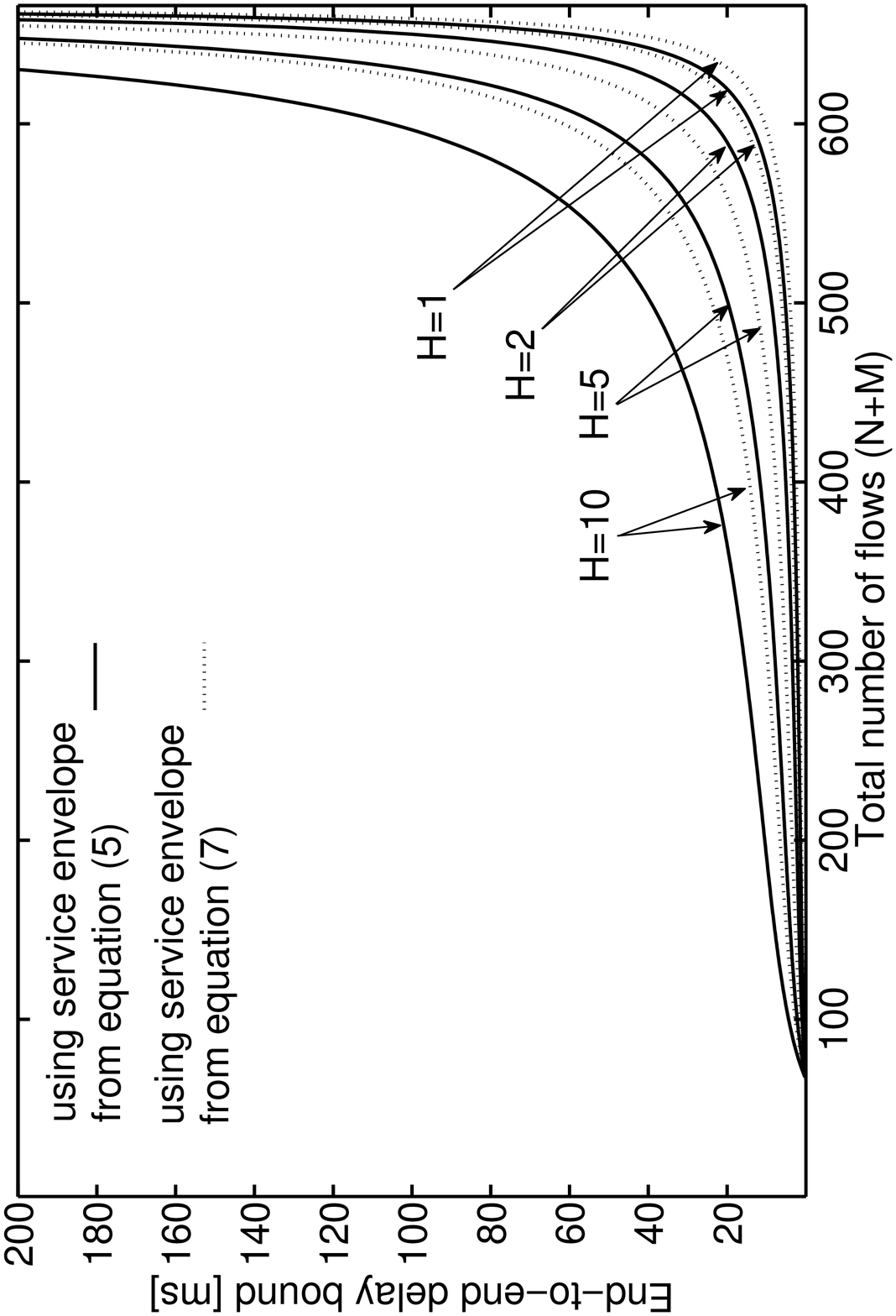}
\label{fig:plot4}
}
\caption{Comparison of End-to-end statistical delay bounds with a violation probability $\varepsilon = 10^{-9}$ for MMOO traffic computed using two different definition of statistical network envelopes}
\label{fig:plot}
\end{figure}

In Fig. \ref{fig:plot} we plot the probabilistic end-to-end delay bound for $N$ through flows in a network with $H=1,2,5,10$ hops for increasing $N+M$ number of flows at each hop while maintaining $N=M$. It can be observed that the new definition of statistical service envelope from equation (\ref{nreffsenv}) together with Theorem \ref{theorem:pbnr} yield a tighter delay bound even for single hop case. The benefit of the new definition of statistical network service envelope is more obvious when the number of nodes $H$ traversed by the through flows is increased. 

\section{Conclusion}
\label{sec:conclusion}
We presented a new formulation of statistical service envelope using the stochastic service process describing the service offered at the network node. We showed for Markov modulated on-off traffic model that the new formulation of statistical service envelope yields end-to-end probabilistic performance measures are tighter than the ones computed using existing state-of-the-art definition of statistical network service envelope and are bounded by ${\cal}(H \log{H})$ for more general $(\sigma(\theta), \rho(\theta))$ - constrained traffic model, where $H$ is the number of nodes traversed by the arrival traffic.

%% The Appendices part is started with the command \appendix;
%% appendix sections are then done as normal sections
%% \appendix

%% \section{}
%% \label{}

%% References
%%
%% Following citation commands can be used in the body text:
%% Usage of \cite is as follows:
%%   \cite{key}         ==>>  [#]
%%   \cite[chap. 2]{key} ==>> [#, chap. 2]
%%

%% References with bibTeX database:

\bibliographystyle{elsarticle-num}
\bibliography{biblio}

\begin{thebibliography}{10}
\expandafter\ifx\csname url\endcsname\relax
  \def\url#1{\texttt{#1}}\fi
\expandafter\ifx\csname urlprefix\endcsname\relax\def\urlprefix{URL }\fi
\expandafter\ifx\csname href\endcsname\relax
  \def\href#1#2{#2} \def\path#1{#1}\fi

\bibitem{boudec:2001}
J.-Y.~L. Boudec, P.~Thiran, Network Calculus: A Theory of Deterministic Queuing
  Systems for the Internet, Springer-Verlag, 2001.

\bibitem{li:2007}
C.~Li, A.~Burchard, J.~Liebeherr, A network calculus with effective bandwidth,
  IEEE/ACM Transactions on Networking 15(6) (2007) 1442--1453.

\bibitem{florin:2006}
F.~Ciucu, A.~Burchard, J.~Liebeherr, Scaling properties of statistical
  end-to-end bounds in the network calculus, Information Theory, IEEE
  Transactions on 52~(6) (2006) 2300 -- 2312.

\bibitem{chang:2000}
C.-S. Chang, Performance Guarantees in Communication Networks, Springer-Verlag,
  2000.

\bibitem{cruz:1996-1}
R.~L. Cruz, Quality of service management in integrated services networks, in:
  Proceedings of 1st Semi-Annual Research Review, June 1996.

\bibitem{yuming:2006}
Y.~Jiang, A basic stochastic network calculus, in: Proceedings of ACM SIGCOMM,
  2006, pp. 123--134.

\bibitem{fidler:2006}
M.~Fidler, An end-to-end probabilistic network calculus with moment generating
  functions, in: Proceedings of IWQoS, 2006.

\bibitem{kelly:1996}
F.~P. Kelly, Notes on effective bandwidths, Stochastic Networks: Theory and
  Applications Oxford, Royal Statistical Society Lecture Notes Series, (1996)
  141--168.

\bibitem{voip_itu:1993}
ITU-T Recommendation P.59, Artificial conversational speech (1993).

\bibitem{maglaris:1988}
B.~Maglaris, D.~Anastassiou, P.~Sen, G.~Karlsson, J.~D. Robbins, Performance
  models of statistical multiplexing in packet video communications, IEEE
  Transactions on Communications 36 (1988) 834--843.

\end{thebibliography}

%% Authors are advised to submit their bibtex database files. They are
%% requested to list a bibtex style file in the manuscript if they do
%% not want to use elsarticle-num.bst.

%% References without bibTeX database:

% \begin{thebibliography}{00}

%% \bibitem must have the following form:
%%   \bibitem{key}...
%%

% \bibitem{}

% \end{thebibliography}

\end{document}